# Measuring and Improving Information Systems Agility Through the Balanced Scorecard Approach

Yassine Rdiouat[1], Samir Bahsani[2], Mouhsine Lakhdissi[3] and Alami Semma[4]

[1] Department of Mathematics and Computer Science, Faculty of Science and Techniques, Hassan 1 University, Settat, Morocco

[2] Department of Mathematics and Computer Science, Faculty of Science and Techniques, Hassan 1 University, Settat, Morocco

[3] Department of Mathematics and Computer Science, Faculty of Science and Techniques, Hassan 1 University, Settat, Morocco

[4] Department of Mathematics and Computer Science, Faculty of Science and Techniques, Hassan 1 University, Settat, Morocco

**Abstract**

Facing an environment increasingly complex, uncertain and changing, even in crisis, organizations are driven to be agile in order to survive. Agility, at the core heart of business strategy, represents the ability to grow in a competitive environment of continuous and unpredictable changes with information systems perceived as one of its main enablers. In other words, to be agile, organizations must be able to rely on agile enterprise information systems/information technology (IT/IS). Since, the agility needs are not the same among stakeholders, the objective of this research is to develop a conceptual model for the achievement and assessment of IT/IS agility from balanced perspectives to support agile organizations. Several researches have indicated that the IT balanced scorecard (BSC) approach is an appropriate technique for evaluating IT performance. This paper provides a balanced-scorecard based framework to evaluate the IS agility through four perspectives: business contribution, user orientation, operation excellence and innovation and competitiveness. The proposed framework, called *IS Agility BSC*, propose a three layer structure for each of the four perspectives: mission, key success factors, and agility evaluation criteria. According to this conceptual model, enterprise information systems agility is measured according to 14 agility key success factors, over the four BSC perspectives, using 42 agility evaluation criteria that are identified based on literature survey methodology.

This paper explores agility in the broader context of the enterprise information systems. The findings will provide, for both researchers and practitioners, a practical approach for achieving and measuring IS agility performance to support organizations in attempt to become agile as a new condition of surviving in the new business world.

## 1. Introduction

Enterprise information systems have become the backbone of organizations, supporting most of their business operations and procedures and often aligned with their business strategy. Since information technology/systems get more complex, enabling new transformations and new forms of work never experienced before such as big data, cloud computing, mobile applications or digital enterprise, quick developments in information systems assessment are crucial to support the company's competitiveness [82].

The increasing unpredictable, dynamic and turbulent environments stress the need for organizations to implement agility as a strategic approach to support the changing business conditions [85].

The concept of agility is being promoted as the solution to give organizations the ability of surviving in a competitive environment of continuous and unpredictable change by reacting quickly and effectively to changing markets, driven by customer-defined products and services [35].

Enterprise information systems are regarded as enablers and facilitators for organizations to be agile [14][59][36][67], making imperative the definition of an adequate framework for the achievement and the assessment of enterprise information systems agility. Many researches about IS agility concept have been conducted at the meantime. However, there remains no common understanding of defining the information systems to support the concept of agility, so a deep understanding of the concept is required to design assessment models of information systems that satisfy the changing needs of organizations.

Most of previous literature explored the IS agility from specific IS area like IT infrastructure agility (e.g. [64]; [106]; [3]), or agile software engineering and agile project management [1][75]. Organizations seeking agility through IS need to consider IS in a broader context as an integrated system for a better fusion between IT and business. The effective contribution of IS, in agile context, cannot be achieved without a better meeting of the balanced expectations of all its stakeholders, this will result in the IS ability to (1) continuously innovate, maintain the competitiveness in the changing environment (2) mobilize internal processes and structures to take advantage of future opportunities (3) satisfy and maintain an agile relationship with users (4) generate IT business value with quicker speed to market.

Measuring IT agility performance should be a key concern of business and IT executives as it supports the justification of future IT investments and demonstrates the effectiveness and added business value of IT. Most of the literature addresses the IS agility measurement in silos without a comprehensive evaluation framework of Agile information systems (AIS) as a general function within the organization. To develop such a comprehensive framework, IT balanced scorecard (IT BSC) is a practical methodology, known by both researcher and practitioners, as performance measurement system. This approach is adopted in this paper to evaluate IS agility in four perspectives: business contribution, user orientation, operation excellence and innovation and competitiveness. In other hand, the literature survey methodology is adopted to identify IT agility key success factors (BSC objectives) and their evaluation criteria (BSC metrics) for each scorecard perspective. Combining these two methodologies helps (i) to give guidelines for achieving IT agility, and (ii) creates a solid foundation for an assessment framework of IT Agility in the organization, called *IS Agility BSC*. The objective is not to concentrate on a on a specific IS area, but to evaluate agility in the broader context of the IS as a critical function in the whole organization.

To set the context, IS agility and balanced scorecard concepts will first be discussed. After that, a balanced scorecard will be introduced as a performance measurement system for IS agility supporting agile organizations to improve their business strategies.

## 2. Research methodology

The Balanced Scorecard is a popular methodology for effective performance measurement that can be applied to IT [95]. Building on IT BSC framework, this approach is adopted in this paper to evaluate IS agility in four perspectives: business contribution, user orientation, operation excellence and innovation and competitiveness. In other hand, the literature survey methodology is adopted to identify IT agility key success factors (BSC objectives) and their evaluation criteria (BSC metrics) for each scorecard perspective. Regarding Rockart [75b], the key success factors, also called critical success factor (CSF), are defined as "The limited number of areas in which results, if they are satisfactory, will ensure successful competitive performance for the organization. Critical Success Factors are strongly related to the mission and strategic goals of your business or project. Whereas the mission and goals focus on the aims and what is to be achieved, Critical Success Factors focus on the most important areas and get to the very heart of both what is to be achieved and how you will achieve it.

The research methodology adopted in this paper is the combination of IT BSC and literature review approaches.
The objective is to build a scorecard for IS agility based on the following steps, as showed in (Fig.1):

1. Literature survey on both IT agility measurement and BSC framework applications
2. Identification of IT BSC as an approach for IT agility assessment
3. Identification of the four BSC perspectives
4. Definition of the mission of each perspective
5. For each mission, identifying key success factors by answering the question "what internal or external area of organization is essential to achieve this mission?"
6. Purification of key success factors by eliminating those judged less important or synonymous. As we identify and evaluate candidate CSF, we may uncover some new strategic area of IT agility. So we may need to review mission and CSFs iteratively.
7. Identification of quantitative and qualitative criteria, based on the literature review of the most relevant agility attributes, to evaluate FCS. The determination of new criteria may lead to review the corresponding CSF and criteria iteratively.

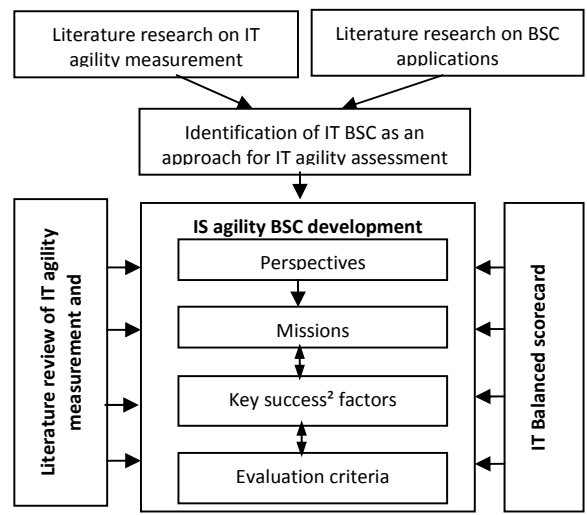

Fig. 1 Research methodology

## 3. The concept of agility

The term of "agility" gained wider recognition since the publication of the Iacocca Institute (Lehigh University, USA) report's entitled *21st Century Manufacturing Enterprise Strategy* [32]. In this report, agility was described as a new industrial order for competitiveness in a volatile manufacturing marketplace. Kidd [45] defined agility as a rapid and proactive adaptation of enterprise elements to unexpected and unpredicted changes. Goldman et al. [31] proposed that agility is the successful application of competitive bases such as speed, flexibility, innovation, and quality by the means of the integration of reconfigurable resources and best practices of knowledge-rich environment to provide customer-driven products and services in a fast changing environment. According to Jackson and Johansson [40] agility is not a goal in itself but the necessary means to maintain the competitiveness in the market characterized by uncertainty and changes. In such environments, companies need to be agile – they need to be able to capitalize on or respond to the opportunities created by new market situations faster than their competitors [31].

The concept of agility has also applied to IS research field. eg. Agile Information Systems Development [4][39][91] or Agile project management [63][25]. Other area are explored, such as, IT infrastructure, IS development, IS organization, and IS personnel as described by Salmela et al [80] in a recent literature review. However, research has shown that neither a widely accepted definition nor commonly used frameworks or concepts exist.

Enterprise information systems are regarded as an enabler for business (enterprise) agility achievement. According to Lui and Piccoli [59] agile information systems enables the firm to identify needed changes in the information processing functionalities required to succeed in the new environment, and which lends itself to the quick and efficient implementation of the needed changes. Agility on the information systems layer applies when changes in information systems are required, due to (external) agility requirements from business. Indeed, given the huge budgets for IT, the top management wants to ensure that IS agility is a key success factor to accomplish business objectives and sustain the business agility. A recent surveys reveal that the main concerns for top executives are business-IT alignment, business agility, business cost reduction, IT-cost reduction; speed to market in that order [54][55][56][66][5]. An important moderating factor in the relationship between IT capabilities and business agility is business-IT alignment [57][58][77]. It is defined as 'the extent to which the IT mission, objectives, and plans support, and are supported by, the organization's mission, objectives, and plans [79]. This alignment creates an integrated organization in which every function, unit, and person are focused on the organization's competitiveness [102]. As agility is the concern of the whole organization, IT agility has to be achieved and evaluated in alignment with the enterprise strategies and respecting the needs of the stakeholders of information systems.

## 4. IS Agility measurement-related works

Few researchers have contributed approaches for measuring agility. According to Tsourveloudis et al [93], agility metrics are difficult to define in general, mainly due to the multidimensionality and vagueness of the concept of agility itself. The works on IS agility measurement concern basically the evaluation of:

1. Agile enterprise with information systems is a major component and enabler [93] [104] [105] [108] [21] [30] [107]
2. Agile information systems of specific-domains, such as, e-government information systems agility [33]
3. Information systems sub-functions agility, such as, information systems development (ISD) agility [34] or Business Intelligence (BI) systems agility [46].
4. Agile enterprise information systems [50] [59]

**Agile enterprise with IS is a major enabler**

Tsourveloudis et al. [93] proposed a fuzzy logic-based framework to evaluate the agility of manufacturing information systems. In this framework, the agility is evaluated according to the four infrastructures of the manufacturing system: i. production ii. market, iii. people, and iv. information. These infrastructures are combined with their corresponding operational parameters to determine the overall agility of the system. Then, the assessment of agility is based on an approximate reasoning method taking into account the knowledge that is included in the fuzzy IF-THEN rules.

Sakthivel et al. [104] and Vinodh et al. [105] designed an agility assessment model consisting of three levels. The first level consisted of five agility enablers such as management responsibility agility, manufacturing management agility, workforce agility, technology agility and manufacturing strategy agility. The second level includes agility criteria and the third level includes agile attributes. This model was used to assess the agility of a firm using scorer model; validation was done using multi-grade fuzzy approach.

Nicola et al. [21] presented a technique for the strategic management of the chain addressing supply planning and allowing the improvement of the Manufacturing Supply Chain (MSC) agility in terms of ability in reconfiguration to meet performance.

Another different approach was proposed by Yauch et al. [108]. They proposed a quantitative index of agility, based on a conceptualization of agility as a performance outcome, which captures both the success of an organization and the turbulence of its business environment. This model integrates operational measures and external parameters to determine the agility of the any type of manufacturing organization.

Ganguly et al. [30] defined three metrics to measure responsiveness, market share and cost effectiveness which would help in measuring a company's agility. They proposed the use of their method along with fuzzy logic approach proposed by Yang et al. [107] in order to arrive at a conclusion regarding the level of agility of any corporate enterprise.

**Agile information systems of specific-domains:**
Gong et al. [33] proposed four principles for creating agility in e-government information systems -particularly in BPM (Business Process Management) systems: i. formulating the business process using business services, ii. integrating and orchestrating business services, iii. separating process, knowledge and resource; and iv. implementing policy by collaboration. Then, based on scenarios derived from the case study, the authors evaluate the level of agility using a set of quantitative and qualitative measures that are defined for each one of the four principles.

**Information systems sub-functions agility:**
In their prospects of a quantitative measurement of agility, Gren et al. [34] conducted a study to validate an agile maturity measurement model of information systems development (ISD) based on statistical tests and empirical data. In this study, they selected to focus on the Sidky's Agile Adoption Framework [88] which is divided into agile levels, principles, practices and concepts, and indicators. First, a pretest was conducted, in this work, as a case study including a survey and focus group. Second, the main study was conducted with 45 employees from two SAP customers in the US. They used internal consistency (by a Cronbach's alpha) as the main measure for reliability and analyzed construct validity by exploratory principal factor analysis (PFA).

Knabke et al. [46] identified agile criteria from IS literature that can be applied to BI system as a major component of information system. They proposed a framework of agility properties categorized into dimensions to provide a valid foundation to evaluate whether a BI system is agile or not. The dimensions obtained are: change behavior, perceived customer value, time, process, model, approach, technology and environment. Each dimension is subdivided into attributes derived from the relevant literature.

**Agile enterprise information systems:**
Lui and Piccoli [59] proposed a framework to evaluate the agility of information systems from the socio-technical perspective. The information system is considered as composed of two sub-systems: a technical system and a social system. The technical sub-system encompasses both technology and process. The social sub-system encompasses the people who are directly involved in the IS and reporting the structure in which, these people are embedded. To measure the information system agility using the socio-technical perspective, Lui and Piccoli used the agility of the four components: i. technology agility, ii. process agility, iii. people agility, and iv. structure agility. The authors argued that, the overall agility of the system is not a simple summing of the obtained scores of agility in these four components, but it depends on their nonlinear relationships. To this end, the authors used the fuzzy logic membership functions to evaluate agility.

Kumar et al. [50] developed an empirically derived framework for better understanding and managing IS flexibility using grounded theory and content analysis. A process model for managing flexibility is presented; it includes steps for understanding contextual factors, recognizing reasons why flexibility is important, evaluating what needs to be flexible, identifying flexibility categories and stakeholders, diagnosing types of flexibility needed, understanding synergies and tradeoffs between them, and prescribing strategies for proactively managing IS flexibility. Three major flexibility categories, flexibility in IS operations, flexibility in IS systems & services development and deployment, and flexibility in IS management, containing 10 IS flexibility types are identified and described.

Although all these works are important, they are either theoretical or address partially the IS agility. Moreover, the agility needs are not the same among IS stakeholders; this aspect needs further investigations supporting a broader range of decision making on evaluating and improving IS agility. Our approach focuses on studying IS agility in an integrated manner considering different aspects of agility in order to meet and balance stockholders expectations using BSC approach. We do not limit ourselves to the IT infrastructure, or software systems, but explore agility in the context of enterprise information systems as a critical function within the whole organization for a better fusion between IT and business.

## 5. Balanced Scorecard Approach

The BSC is a performance measurement approach, introduced at the enterprise level [42] since 1992 by Kaplan and Norton, it allows managers to look at their business performance from four performance perspectives (financial, internal processes, customer and innovation and learning). The fundamental premise of the BSC framework is to integrate all the interests of key stakeholders (e.g., executives, IT managers, Business unit's managers, customers, employees, etc.) on a scorecard. The term "balanced" reflects the balance provided between short-and long-term objectives, between quantitative and qualitative performance measures, and between different perspectives. For this balanced measurement framework, Kaplan and Norton proposed a three layer structure for each of the four perspectives: mission, objectives, and measures. Each perspective can be explained by a key question with which it is associated. The answers to each key question become the objectives associated with that perspective, and performance is then judged by the progress to achieving these objectives [12]. There is an explicit causal relationship between the perspectives: good performance in the innovation and learning objectives generally drives improvements in the internal business process objectives, which should improve the organization in the eyes of the customer, which ultimately leads to improved financial results. According to Van Grembergen & De Haes [22] each of these perspectives has to be translated into the corresponding metrics and measures that assess the current situation. These assessments should be repeated periodically and have to be confronted with the objectives that have to be fixed in advance [53].

BSC can be applied to IT as initially described by Van Grembergen et al. [98] [98] [100]. The adaptations made by these authors generated a generic scorecard for IT known in the literature as IT BSC. The success of the BSC in IT is due to its flexible design and comprehensive nature. By adding or altering individual measures and perspectives, the BSC can be tailored to suit the strategy of any organization [28]. Moreover, a cascade of BSCs can be used across multiple organizational levels supporting strategic alignment [96] [65].

Due to its easy adaptability, several issues have been addressed by using the IT BSC framework, such as, evaluating IT projects [8], evaluate IT departments performance [51], evaluating Information Systems (IS) performance [76], prevent sub-optimization of IT performance [7], ensuring strategic alignment between IT and the business [97], integration of business and IT governance [10], and others.

## 6. Developing an IT Agile Balanced Scorecard

It was demonstrated previously that the BSC concept can be applied to IT. This paper suggests that the BSC has the potential to help organizations evaluate their IT/IS Agility, in a holistic manner, through the process of measuring and analyzing induced performance improvement.

To apply this approach to the IT Agility, the four perspectives of the generic IT BSC should be changed accordingly. In (Fig. 2), a generic IT BSC is shown [96].

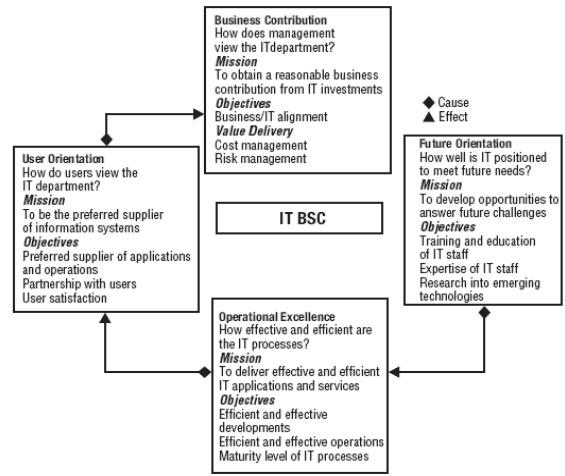

Fig. 2 Generic IT Balanced Scorecard.

The user orientation perspective represents the user evaluation of IT. The operational excellence perspective represents the IT processes employed to develop and deliver the applications. The future orientation perspective represents the human and technology resources needed by IT to deliver its services over time. The business contribution perspective captures the business value created from the IT investments.

Based on the literature review and according to the generic IT BSC, the four perspectives of IT/IS agility BSC are built (Fig. 3).

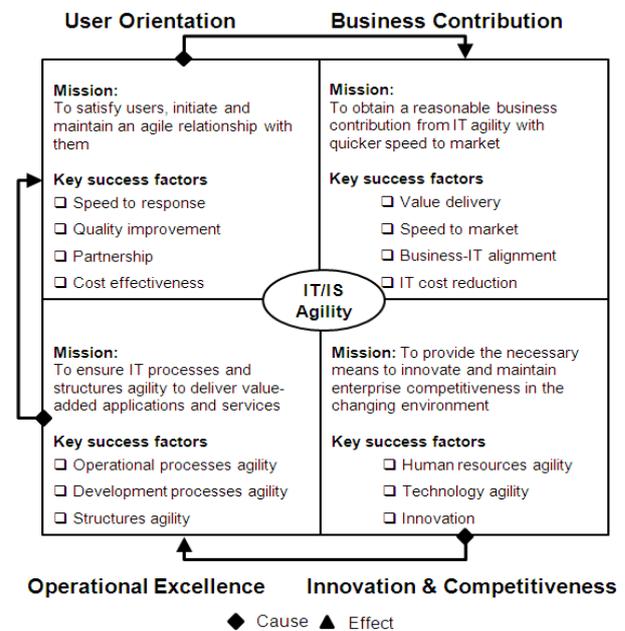

Fig. 3 IT/IS Agility scorecard perspectives and their cause-and-effect Relationships

*Business contribution* perspective evaluates the IT agility from the viewpoints of executive management. It captures the business value created by the IT agility, as enabler of business agility, and investigates the link between IT agile implementation and associated tangible and intangible benefits experienced by the organization [89]. Moreover, different studies conducted to explore if and how IT capabilities contribute to higher levels of business (Or enterprise) agility performance [81][106][15]. Cap Gemini [17] has found in a survey among 300 CIOs and other IT executives worldwide that all organizations with high perceived business agility also scored high on IT/IS agility, and 93% of CIOs agree that business value is created by IT agility. Hence, we define the mission of this perspective as the ability *to obtain a reasonable business contribution from IT agility with quicker speed to market.*

The IT BSC Framework evaluates user orientation perspective from the viewpoints of internal business users, however, regarding the emergence of digital links between the company and its customers including e-business and social networks, the technical objects that provides IT department become directly visible to the customer. Hence, the IS becomes a direct actor in customer relationship. This involves a new requirement of the IS function on the customer experience, the ergonomics and the quality of service. Therefore, the *user orientation* perspective of IS agility BSC evaluates the IT agility from a broader viewpoints of IS users including customer (end user), partners, in addition to internal business users. It answers the question "How do stakeholders view the agility of the IT department?" This perspective is based largely on the IS ability to initiate and maintain an agile relationship with its clients and partners, including internal business users, based on core principles such as, user partnership built on collaboration, sharing, transparency and trust, tailor-made products, cost optimization, service quality, technical support, responsiveness to change and user satisfaction. Then, we can deduce that the mission of this perspective is to "*Satisfy users, initiate and maintain an agile relationship with them.*"

The *operational excellence* perspective answers the key question "How agile are the processes and structures of IT department to deliver applications and services in order to satisfy the stockholders?" Therefore, the mission is *"To ensure IT processes and structures agility to deliver value-added applications and services"*

The *innovation and competitiveness* perspective is adapted from the 'innovation and learning' perspective of the BSC. It shows the agility of IT from the viewpoint of the IT organization itself, and answers the key question "How well is IT positioned to innovate and provide more competitive advantages?" According to [89], this perspective is focusing on the long-term achievement of the organization goals and how the newly implemented technology creates competitive advantages in the future e.g., potential for global co-operation, enhancing organizational image, and attracting more sophisticated clients. This perspective is perhaps the most difficult to quantify but has the greatest potential to provide the necessary means to preserve IS agility. Competitiveness is representative of management's ability to instill the necessary cultural change to embrace innovative technology. Employees with the ability to adapt to an ever-changing work environment will be more receptive to new IT/IS applications, which improve operational efficiency. Hence, the mission is *"To provide the necessary means to innovate and maintain the enterprise competitiveness in the changing environment"*

A cause-and-effect relationship must be defined throughout the whole scorecard. As shown in (Fig. 3), the four perspectives have amongst each other cause-and-effect relationships. Over all, effective agility drivers like better education of IT staff (innovation and competitiveness) is an enabler (performance driver) for a better and faster developed applications (operational excellence perspective) that in turn is an enabler for measuring up user expectations (user orientation perspective) that eventually will enhance higher business value of IT and more business agility (business contribution).

Consequently, IT agility can be achieved through a continuous improvement approach, based on the ability of AIS to (1) continuously innovate, maintain the competitiveness in the changing environment (2) mobilize internal processes and structures to take advantage of future opportunities (3) satisfy and maintain an agile relationship with users (4) generate IT business value with quicker speed to market.

## 7. Key success factors and evaluation criteria for IS Agility BSC

The objectives of the IS Agility BSC represents the key success factors of IT agility to achieve the mission of each perspective. These objectives are built based on IT BSC framework and literature review as showed in (Table 1). The metrics refers to the evaluation criteria based on the objectives can be assessed. This paragraph discusses furthermore both key success factors and evaluation criteria of the proposed IS Agility BSC framework.

Table 1: IT/IS scorecard key success factors

| Perspectives | Key success factors | References |
|---|---|---|
| Business contribution | Value delivery | Adapted from IT BSC [95][96] |
| | Speed to market | [55] [56] [66][5] |
| | Business-IT alignment | Adapted from IT BSC [95][96] |

| | IT cost reduction | [55] [56] [66][5] Adapted from IT BSC [95][96] |
|---|---|---|
| User orientation | Speed to response | [2] [83] |
| | Quality improvement | [2] [109] |
| | Partnership with users | [9] [57] [58] [70] [103] |
| | Cost effectiveness | [2] [81] [85] |
| Operational excellence | Operational processes agility | [109] Adapted from IT BSC [95] |
| | Development processes agility | Adapted from IT BSC [95] |
| | Structures agility | [109]; Adapted from IT BSC [95][96] |
| Innovation and competitiveness | Human resources agility | [59][109] |
| | Technology agility | |
| | Innovation | [81][109] |

**Business contribution perspective:**
The main concern of the agility is the strategic business-IT alignment. It is logic that the IS Agility BSC starts with the business contribution perspective. The IS Agility BSC showed in (Fig. 3 and Table 1) links with business through the business contribution perspective (business/IT alignment, value delivery, IT cost reduction and Speed to Market). The main measurement challenge is within the area of *strategic alignment*. Some criteria are used by Grembergen [95]:
- Business goals supported by IT goals
- Operational plan/budget approval

The *value delivery* means delivering value and solutions to the customer, rather than products, in order to bring product to the market as rapidly as possible [28]. It can be directly measured against the objectives of the overall business. Traditional financial evaluations can be used for the assessment of the value delivery, such as the return on investment (ROI), net present value (NPV), internal rate of return (IRR) and payback period (PB). On another side, the business value can be created with innovation when it is reflected in high value-added products and services. Indeed, regarding Jack Welch, CEO from General Electric, "If the rate of change on the outside exceeds the rate of change on the inside, then the end is near". Moreover, in Schumpeter's theory, innovation is the source of value creation. Schumpeterian innovation emphasizes the importance of technology and considers novel combinations of resources (and the services they provide) as the foundations of new products and production methods. These, in turn, lead to the transformation of markets and industries, hence to economic development [7b]. Thus, in this paper, we introduced "Business value based on IT innovation" as a new criterion to evaluate the value created by IT based on the exploitations of new business opportunities generated by IT innovation witch is in the heart of IT agility.

*IT cost reduction* is a traditional financial objective and is measured through the attainment of expense and recovery targets. The expenses refer to the costs that the IT organization has made for the business, and the recovery refers to the allocation of costs to IT services and the internal charge back to the business [96].

*Speed to market* is vital for business growth in today's competitive economy, especially as organizations continue to increase their use of IT for competitive advantage [54]. It is defined as the time it takes to recognize a market opportunity, translate this into a product or service and bring it to the market [18]. This can be assessed trough the "Time to Market of new products or services introduction" [85]. Other objectives related to the same concepts such as delivery speed [2] remain in the operational level. These will be discussed in the user orientation perspective.

Table 2: Evaluation criteria of business contribution perspective

| Key factors | Criteria | References |
|---|---|---|
| Value delivery | Business value of major IT projects | Adapted from IT BSC [95][96] |
| | Business value based on IT innovation | New |
| Speed to market | New products introduction vs. competition | Adapted from [110] [42] |
| Strategic business-IT alignment | Business goals supported by IT goals | Adapted from IT BSC [95][96] |
| | Operational plan/budget approval | |
| IT cost reduction | Attainment of expense and recovery targets | |

**User orientation perspective:**
The challenge for IT department is how to serve and satisfy the maximum our users? This perspective provides answers to the key questions of these stakeholders concerning IT agility to meet business expectations, and maintain an agile relationship with customers and partners. Regarding Norton and Kaplan [42], customer's concerns, at the enterprise level, tend to fall into four categories: time, quality, performance and service, and cost. In the same way, Agarwal [2] formulated more detailed agile attributes (customer satisfaction, quality improvement, cost minimization, delivery speed, new product introduction, service level improvement, lead-time reduction). As shown in (Table 3), the issues this perspective focuses on are the basic performance objectives, which the user expects, of speed to response, quality improvement, and partnership with users. *Speed to response* or Time agility [83] is the capability of an organization to rapidly execute decision making, operational cycles and reconfiguration of corporate structures [68]. At the heart of agility concept, many researchers consider the time and speed dimension as an intrinsic concept for agility achievement. In this paper we consider *Speed to response as the time it takes to respond to user needs* (time it takes to response to customer demand, time to deliver new products…). Nevertheless, Time is also synonym of quickness [85] which is on the main agility capabilities that will be used as evaluation criteria through the IT agility scorecard. Speed to response is relevant and can be assessed by various criteria, such as:
- *Delivery speed*
- *Speed to decision making*
- *Speed of data access*

In the *quality improvement* area, several criteria can be used, such as:
- *Data transparency:* represents the level of data quality and availability to IS users. It refers, moreover, to the level of alignment between business needs and IT needs for data business intelligence [71].
- *Products value addition*
- *Quality over product life*
- *User satisfaction*

A major concern of business is the level of IT *costs effectiveness*, which can be measured through criteria, such as, attainment of unit cost targets and IT costs charged back to the business unit [96]. *Partnership* represents the level of alignment and relationship agility between IT department and its users. This includes:
- *Collaboration:* represents the level of cooperation between IT department and its users to enhance competitiveness and allows responding to change than following a plan [11] [26].
- *Communication efficiency:* refers to the level of internal and external communication management within the organization for faster decision-making [58]
- *Trust based relationships:* determines the level of trust between the IT and its users.

Table 3: Evaluation criteria of user orientation perspective

| Key factors | Criteria | References |
|---|---|---|
| Speed to response | Delivery speed | [2] |
| | Speed of data access | [11] |
| | Speed to decision making | [27] |
| Quality improvement | Data transparency | [71] |
| | Products value addition | [109] |
| | Quality over product life | |
| | User satisfaction | |
| Partnership | Collaboration | [103] [9][11] [86] [69] |
| | Communications efficiency | [36] |
| | Trust based relationships | [20] [73] [101] [38] |
| Cost effectiveness | Attainment of unit cost targets | Adapted from IT BSC [95] [96] |
| | IT costs charged back to the business unit | |

**Operational Excellence perspective:**
As discussed in the previous paragraphs, operational excellence perspective represents the agility of IT processes and structures employed to develop and deliver IT services and applications. This can mean offering a wider variety of products and services in response to changes, increasing customization and developing new products and services. Moreover, information systems must be able to make these changes quickly and at low cost; it must therefore design agility into its operations and services. In other words, IT department should enable the operational agility to accomplish speed, accuracy, and cost economy in the exploitation of innovation opportunities [81]. The issues that are of focus here, as displayed in (Table 1), are the agility performance of operational process, development process and structures. The operational process agility area answers key questions like productivity, reliability of IT process and can be assessed through several evaluation criteria as shown in (Table 4):
- *Integrability*: is mainly concerned with the integration of IT/IS into the organization and the role it plays in process co-ordination and integration of internal IS components [89] [109].
- *Responsiveness to change:* is the process ability to respond to changing conditions and customer interactions as they occur.
- *Flexibility:* refers to the level of simplification and the capability to rapidly change from one task or from one production route to another, including the ability to change from one situation to another [93].
- *Setup times/costs:* represents the level of process re-configurability. According to Lui and Piccoli [59], this criteria deals with the time and costs of setting up alternative processes in the face of changes in demand, market conditions, strategy, etc. For example, a process is considered more agile when it can produce different types of products and services without major setup time and cost as demand changes.
- *Performance evaluation of IT processes:* deals with process governance area of IT processes.

The *development process agility* area refers to measures like level of productivity and quality of delivery from clients developing their requirements for new capital projects through the design to the final delivery and maintenance of new products and services. This can be assessed through criteria like:
- *Customized products and services*
- *Short development cycle time*
- *First time right designs*
- *Continuous integration*

*Structures agility* refers to "the degree of flexibility and decision-making ability afforded to individual members of the information system [59]. It can be evaluated trough criteria, such as, leaderships or workforce empowerment, distributed decision-making authority, and flatter managerial hierarchies [93]. An empowered workforce, and distributed decision-making authority allow employees to take leadership in decision making and to make it quickly. Flatter managerial hierarchies enhance communication within the organization and speed up the decision-making process in the face of more general and strategic level changes. Eshlaghy et al. [29] identified 12 factors that have an effect on organizational agility by applying path analysis. Interestingly, the most significant are leadership, and organization commitment. Organizational commitment refers to the extent to which the employees of an organization see themselves as belonging to the organization (or parts of it) and feel attached to it [60] [94], a questionnaire in six language is established by Kanning [41] for the validation of the organizational commitment. Hence, as shown in (Table 4), the selected evaluation criteria for this key success factor are:
- *Leadership*
- *Organization commitment*
- *Flatter managerial hierarchies*

Table 4: Evaluation criteria of operational excellence perspective

| Key factors | Criteria | References |
|---|---|---|
| Operational processes agility | Integrability | [2][85][18][13][19][20][73] |
| | Responsiveness to change | [81[90] |
| | Flexibility | [31] [43] [44] |
| | Setup times/costs | [23] [87] |
| | Performance evaluation of IT processes | Adapted from IT BSC [95]| |
| Development process agility | Customized products and services | [35] [45] [72] [93] |
| | Short development cycle time | [109] |
| | First time right designs | |
| | Continuous integration | [105b] |
| Structures agility | Leadership | [29] |
| | Organization commitment | |
| | Flatter managerial hierarchies | [29] [93] |

**Innovation and competiveness:**
It evaluates the agility from the viewpoints of IT organization itself. The issues focused on, as depicted in (Table 1), are Human resources agility, Technology agility and Innovation. These three components are also the main agility providers as described by Sharifi and Zhang [85].
*Human resources agility* is the degree to which individuals, associated with the information systems, possess knowledge and skills that are both varied and easily redeployable in the face of change [59]. It can be measured through criteria like training level and education, Competency and adaptability as shown in (Table 5). According to Goldman et al. [31], an agile competitive environment is where the people skills, knowledge, and experience are the main differentiators between the companies.

*Technology agility* area represents the degree of flexibility of the information technology and the extent to which the IT components of the information system lends itself to rapid adjustment when needed [59]. Some criteria can be applied as showed in (Table 5).

*Innovation* is the successful exploration of new ideas for products, services, procedures [84]. The main focus of the IT department is to provide innovative technology to enable development of new products and services according the business orientations. Gurd [37] introduced on his Agile BSC two criteria related to innovation assessment, product innovation and process innovation which represent the rate of improvements of existing products and processes or the introduction of new ones. Moreover, in their study on the IT impact on business innovation, Aubert et al. [8b] distinguish six types of innovations: commercial innovation, organizational innovation, technological innovation, products innovation, processes innovation and business model innovations. However, innovation on purely IT aspects cannot be finality in itself for IT department without generating business value for the whole company. Especially as today's information system exceeding its traditional boundaries to position itself as a partner and a company's growth driver. Thus, IT innovation, even irreproachable, cannot be the unique purpose of IT teams. It only makes sense by its actual use for business innovation, in other words, when the IT becomes one of the innovation sources, among many others, positively impacting the business with its different forms. We can deduce through this work a new IT agility evaluation criterion called "Innovations with business value addition ". Some evaluation criteria are shown in (Table 5) to evaluate Innovation key success factor.

Table 5: Evaluation criteria of innovation and competitiveness perspective

| Key factors | Criteria | References |
|---|---|---|
| Human resources agility | Continuous training and development | [109] |
| | Competency | [6] [74] [52] [62] |
| | Adaptability | |
| Technology agility | Standardization | [24] [15] [78] [49] [92] |
| | Connectivity | |
| | Compatibility | |
| | Modularity | |
| | Scalability | [48] [23] |
| | Reconfigurability | |
| Innovation | Technology awareness | [37] |
| | Customer-driven innovations | |
| | Innovations with business value addition | New |

## 8. Map for IS Agility BSC

In order to ensure IS agility improvement, as a strategic objective of the organization, our BSC, to be well-developed, should have a card called "strategic map" to evaluate and make visually explicit perspectives, objectives and measures, and the main links between the different perspectives of BSC through reports of cause and effect as showed in (Fig. 3). An example of such cause-effect is that the organization aims to sustainably improve the Speed-to-Market (business contribution perspective). To achieve this goal, the level of partnership and synergy between business teams and IT team should rise, and all business demands must be processed in time (user orientation perspective). This also means that the agility of operational processes and development processes should be improved in order to respond quickly to changes, and the internal structures must enhance more leadership to accelerate the decision making process on one hand, and on the other, the organization of the IT department is agile enough for better collaboration internally and with business users. The objectives of operational excellence level can only be achieved if the information system permanently improves its capacity for innovation and the level of its technological advance, and whether the sense of employee's agility performs by learning, competency and adaptability (perspective innovation and competitiveness).

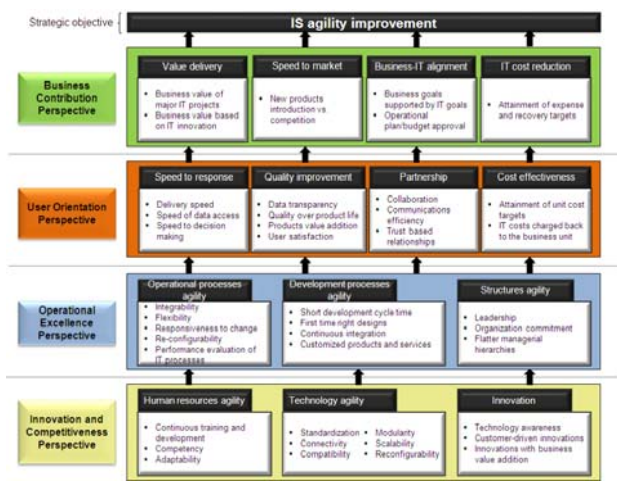

Fig. 3 Strategic map for IT/IS Agility scorecard

## 9. Conclusion and future work

Organizations are driven to be agile in order to survive facing the business changing environments. Enterprise information systems are regarded as the main enabler of the organization agility in order to maintain competitive advantages. Therefore, the achievement and evaluation of IS agility is a strategic approach not only for the IT department but for the whole organization. An important conclusion of the paper is that IS agility can be achieved through a continuous improvement approach, based on the ability of AIS to (1) continuously innovate, maintain the competitiveness in the changing environment (2) mobilize internal processes and structures to take advantage of future opportunities (3) satisfy and maintain an agile relationship with users (4) generate IT business value with quicker speed to market.. In this context, a framework called *IS Agility BSC* is developed based on IT BSC approach and the literature review on previous works on IT agility manufacturing and assessment. The conceptual model presented in this paper gives practical guidelines to the achievement and the assessment of information systems agility, through operational objectives with evaluation criteria and over four balanced perspectives: business contribution, user orientation, operational excellence and innovation and competitiveness. The framework presented in this work can be tested with the help of suitable empirical and multiple case studies. As next step, since the evaluation criteria contain both tangible and intangible metrics, we aim to expand the present work by an evaluation methodology, such as a scoring model, in order to determine the IS agility level with the help of appropriate empirical and case study research. This work is intended to contribute to the theory-building process in the fields of agility and information systems.

The following are some of the major issues that could be considered by researchers and practitioners:

- The conceptual model for agility evaluation could also be extended to more key success factors and criteria according to the business context, since no organization is similar to another.
- IS Agility BSC must go beyond the IT department level and must be integrated across the enterprise in order to generate business value. This can be realized through the development of a business agility balanced scorecard to establish a linkage with IT Agility balanced scorecards.
- Cascading IS Agility BSC to IT Units needs to be defined, so, the objectives become more operational and tactical, as do the performance measures. Indeed, individual scorecards should be developed to link day-to-day work with departmental goals and enterprise vision.
- Investigating more the link between IT agility and IT governance since IT agility is rarely measured as part of IT governance.

# References

[1] Abrahamsson, P., Warsta, J., Siponen, M. T., & Ronkainen, J. (2003, May). New directions on agile methods: a comparative analysis. In Software Engineering, 2003. Proceedings. 25th International Conference on (pp. 244-254). Ieee.

[2] Agarwal, A., Shankar, R., & Tiwari, M. K. (2007). Modeling agility of supply chain. Industrial Marketing Management, 36(4), 443-457.

[3] Ahsan, M., Ye-Ngo, L. (2005). The relationship between IT infrastructure and strategic agility in organizations. Paper presented at the Eleventh Americas Conference on Information Systems, Omaha, NE.

[4] Alaa, G., & Fitzgerald, G. (2013). Re-conceptualizing agile information systems development using complex adaptive systems theory. Emergence: Complexity & Organization, 15(3).

[5] Alenezi, H., Tarhini, A. and Masa'deh, R. (2015). Investigating the Strategic Relationship between Information Quality and E-Government Benefits: A Literature Review. International Review of Social Sciences and Humanities, 9 (1), 33-50.

[6] Almahamid, S., Awwad, A., & McAdams, A. C. (2010). Effects of organizational agility and knowledge sharing on competitive advantage: an empirical study in Jordan. International Journal of Management, 27(3), 387.

[7] Amado, C. A., Santos, S. P., & Marques, P. M. (2012). Integrating the Data Envelopment Analysis and the Balanced Scorecard approaches for enhanced performance assessment. Omega, 40(3), 390-403.

[7b] Amit, R., & Zott, C. (2000). Value creation in e-business. INSEAD.

[8] Asosheh, A., Nalchigar, S., & Jamporazmey, M. (2010). Information technology project evaluation: An integrated data envelopment analysis and balanced scorecard approach. Expert Systems with Applications, 37(8), 5931-5938.

[8b] Aubert, B., Cohendet, P., Da Silva, L., Grandadam, D., Guimaron, J., & Montreuil, B. (2010). L'innovation et les technologies de l'Information et des communications. Centre sur la productivité et la prospérité (HEC Montréal), & CEFRIO.

[9] Best, R.J. (2001). Market-Based Management: Strategies for Growing Customer Value and Profitability. Prentice Hall, Third Edition, Upper Saddle River, NJ.

[10] Borousan, E., Hojabri, R., Manafi, M., & Hooman, A. (2011). Balanced Scorecard: A tool for measuring and modifying IT Governance in Healthcare organizations. International Journal of Innovation, Management and Technology, 2(2), 141.

[11] Breu, K., Hemingway, C. J., Strathern, M., Bridger, D. (2002). Workforce agility: the new employee strategy for the knowledge economy. Journal of of Information Technology, 17(1), 21-31.

[12] Bloomfield, C. (2002). Bringing the Balanced Scorecard to Life: The Microsoft Balanced Scorecard Framework. Microsoft Corporation White Paper.

[13] Bruce M., Daly, L., Towers, N. (2004). Lean or agile: A solution for supply chain management in the textiles and clothing industry? International Journal of Operations & Production Management, 24(2), 151-170.

[14] Byrd, T. A., Turner, D.E. (2001). An explorative examination of the relationship between flexible IT infrastructure and competitive advantage. Information & Management,39, 41-52.

[15] Byrd, T. A., Turner, D.E. (2000). Measuring the Flexibility of Information Technology Infrastructure: Exploratory Analysis of a Construct. Journal of Management Information Systems, 17(1), 167-208.

[16] Cao, Q., & Dowlatshahi, S. (2005). The impact of alignment between virtual enterprise and information technology on business performance in an agile manufacturing environment. Journal of Operations Management, 23(5), 531-550.

[17] Cap Gemini. (2007). Global CIO Survey 2007. IT Agility: Enabling Business freedom: Capgemini.

[18] Christopher, M., Lowson, R., Peck, H. (2004). Creating agile supply chains in the fashion industry. International Journal of Retail & Distribution Management, 32(8/9), 367-376.

[19] Christopher, M. (2000). The agile supply chain: competing in volatile markets. Industrial marketing management, 29(1), 37-44.

[20] Christopher, M., Towill, D. R. (2000). Supply chain migration from lean and functional to agile and customised. Supply Chain Management: An International Journal, 5(4), 206-213.

[21] Costantino, N., Dotoli, M., Falagario, M., Fanti, M. P., & Mangini, A. M. (2012). A model for supply management of agile manufacturing supply chains. International Journal of Production Economics, 135(1), 451-457.

[22] De Haes, S., & Van Grembergen, W. (2009). An exploratory study into IT governance implementations and its impact on business/IT alignment. Information Systems Management, 26(2), 123-137.

[23] Dove, R. (2001). Response ability: the language, structure and culture of the agile enterprise: John Wiley & Sons.

[24] Duncan, N. B. (1995). Capturing flexibility of information technology infrastructure: a study of resource characteristics and their measure. Journal of Management Information Systems, 12(2), 37-57.

[25] Dyba, T., & Dingsoyr, T. (2015, May). Agile Project Management: From Self-Managing Teams to Large-Scale Development. In Software Engineering (ICSE), 2015 IEEE/ACM 37th IEEE International Conference on (Vol. 2, pp. 945-946). IEEE.

[26] Dyer, L., & Shafer, R. A. (2003). Dynamic organizations: Achieving marketplace and organizational agility with people.

[27] Ekman, A., Angwin, D. (2007). Industry patterns of agility: a study of the role of Information Systems and Information Technology as an antecedent of strategic agility within European organisations. International Journal of Agile Systems and Management, 2(4), 360-375.

[28] Erek, K. (2011). From green IT to sustainable information systems management: Managing and measuring sustainability in IT organizations. In Proceedings of the European, Mediterranean & Middle Eastern Conference on Information Systems (pp. 1-16).

[29] Eshlaghy, A. T., Mashayekhi, A. N., Rajabzadeh, A., & Razavian, M. M. (2010). Applying path analysis method in defining effective factors in organisation agility. International Journal of Production Research, 48(6), 1765-1786.
[30] Ganguly, A., Nilchiani, R., & Farr, J. V. (2009). Evaluating agility in corporate enterprises. International Journal of Production Economics, 118(2), 410-423.
[31] Goldman, S., Nagel, R., Preiss, K. (1995). Agile Competitors and Virtual Organizations. New York: Van Nostrand Reinhold.
[32] Goldman, S. L. and Preiss, K. (eds): 21st Century Manufacturing Enterprise Strategy: An Industry-Led View, Bethlehem, PA, Iacocca Institute at Lehigh University, 1991.
[33] Gong, Y., & Janssen, M. (2012). From policy implementation to business process management: Principles for creating flexibility and agility. Government Information Quarterly, 29, S61-S71.
[34] Gren, L., Torkar, R., & Feldt, R. (2015). The Prospects of a Quantitative Measurement of Agility: A Validation Study on an Agile Maturity Model. Journal of Systems and Software.
[35] Gunasekaran, A. (1999). Agile manufacturing: a framework for research and development. International journal of production economics, 62(1), 87-105.
[36] Gunasekaran, A. (1998). Agile manufacturing: enablers and an implementation framework. International Journal of Production Research, 36(5), 1223-1247.
[37] Gurd, B., & Ifandoudas, P. (2014). Moving towards agility: the contribution of a modified balanced scorecard system. Measuring Business Excellence, 18(2), 1-13.
[38] Handfield, R. B., Bechtel, C. (2002). The role of trust and relationship structure in improving supply chain responsiveness. Industrial Marketing Management, 31(4), 367-382.
[39] Hummel, M., & Epp, A. (2015, January). Success Factors of Agile Information Systems Development: A Qualitative Study. In System Sciences (HICSS), 2015 48th Hawaii International Conference on (pp. 5045-5054). IEEE.
[40] Jackson, M., Johansson, C., 2003. Agility analysis from a production system perspective. Intergarted Manufacturing Systems 14 (6),482–488.
[41] Kanning, U. P., & Hill, A. (2013). Validation of the Organizational Commitment Questionnaire (OCQ) in six Languages. Journal of Business and Media Psychology, 4, 11-20.
[42] Kaplan, R. S., & Norton, D. P. (1996). The balanced scorecard: translating strategy into action. Harvard Business Press.
[43] Kidd, P. T. (2000). Agile manufacturing enterprise strategy: Cheshirenbury.
[44] Kidd, P. T. (1995). Agile Corporations: Business Enterprises in the 21st Century - An Executive Guide. Cheshire: Henbury.
[45] Kidd, P. T. (1994). Agile Manufacturing: Forging New Frontiers. Wokingham: Addison-Wesley.
[46] Knabke, T., & Olbrich, S. (2013, January). Understanding Information System Agility--The Example of Business Intelligence. In System Sciences (HICSS), 2013 46th Hawaii International Conference on (pp. 3817-3826). IEEE.
[47] Knoll, K., & Jarvenpaa, S. L. (1995, January). Learning virtual team collaboration. In Hawaii International Conference on Systems Sciences Proceedings (Vol. 4, pp. 92-101).
[48] Knoll, K., & Jarvenpaa, S. L. (1994, April). Information technology alignment or "fit" in highly turbulent environments: the concept of flexibility. In Proceedings of the 1994 computer personnel research conference on Reinventing IS: managing information technology in changing organizations: (pp. 1-14).
[49] Konsynski, B., Tiwana, A. (2004). The improvisation-efficiency paradox in inter-firm electronic networks: governance and architecture considerations. Journal of Information Technology, 19(4), 234-243.
[50] Kumar, R. L., & Stylianou, A. C. (2014). A process model for analyzing and managing flexibility in information systems. European Journal of Information Systems, 23(2), 151-184.
[51] Lee, A. H., Chen, W. C., & Chang, C. J. (2008). A fuzzy AHP and BSC approach for evaluating performance of IT department in the manufacturing industry in Taiwan. Expert systems with applications, 34(1), 96-107
[52] Lin, C-T., Chiu, H. and Tseung, Y.H. (2006) «Agility evaluation using fuzzy logic » -International Journal of Production Economics, Vol.101 No.2, pp.353-368.
[53] Lorences, P. P., & Ávila, L. G. (2014). The construction of a scorecard of information technology in a company. Visión de Futuro, 18(2).
[54] Luftman, J., & Derksen, B. (2015). European key IT and management issues & trends for 2015. CIONET Europe and Business & IT Trend Institute.
[55] Luftman, J., & Derksen, B. (2014). European key IT and management issues & trends for 2014. CIONET Europe and Business & IT Trend Institute.
[56] Luftman, J., and Ben-Zvi, T. 2010. "Key Issues for It Executives 2010: Judicious It Investments Continue Post-Recession," MIS Quarterly Executive (9:4), pp. 263-273
[57] Luftman, J. (2003). Assessing IT/business alignment. Information Systems Management, 20(4), 9-15.
[58] Luftman, J. (2000). Assessing IT-business alignment. Communications of AIS, 4(14), 1-50.
[59] Lui, T. W., & Piccoli, G. (2007). Degrees of agility: implications for information systems design and firm strategy. Agile information systems: Conceptualization, construction, and management, 122-133.
[60] Meyer, J. P., Kam, C., Gildenberg, I. & Bremner, N. L. (2013). Organizational commitment in the military: Application of a profile approach. Military Psychology, 25, 381-401.
[61] McCoy, D. W., & Plummer, D. C. (2006). Defining, cultivating and measuring enterprise agility. Gartner research, 28.
[62] McCormack, K. P., & Johnson, W. C. (2001). Business process orientation: gaining the e-business competitive advantage. CRC Press.
[63] Moran, A. (2015). Agile Project Management. In Managing Agile (pp. 71-101). Springer International Publishing.
[64] Nelson, K. M., Cooprider, J. G. (2001). The relationship of software system flexibility to software system and team

[64] performance, Twenty-Second International Conference on Information Systems (pp. 23–32).
[65] Niven Paul, R. (2006). Balanced Scorecard step by step.
[66] Orozco, J., Tarhini, A., Masa'deh, R. and Tarhini, T. (2015). A framework of IS/business alignment management practices to improve the design of IT Governance architectures. International Journal of Business and Management,10(4),1-12.
[67] Overby, E., Bharadwaj, A. and Sambamurthy, V. (2006). Enterprise agility and the enabling role of information technology. European Journal of Information Systems, 15 (2), 120-131.
[68] Palethorpe, D. (2003). The agile enterprise: a foresight scenario sheet: MANTYS research report.
[69] Pimentel Claro, D. and Oliveira Claro, P.B. (2010), "Collaborative buyer – supplier relationships and downstream information in marketing channels", Industrial Marketing Management, Vol. 39 No. 2, pp. 221-228.
[70] Poels, R. (2006). Beïnvloeden en meten van business–IT alignment. Free University of Amsterdam,Amsterdam.
[71] Prahalad, C. K., Krishnan, M.S., Ramaswamy, V. (2002). Manager as Consumer - The essence of Agility (No. Working Paper #02-013): University of Michigan Business School
[72] Preiss, K. G., S.L., Nagel, R.N. (1996). Cooperate to compete: building agile business relationships. New York: Van Nostrand Reinhold.
[73] Power, D. J., Sohal, A. S., Rahman, S. U. (2001). Critical success factors in agile supply chain management. International Journal Physical Distribution and Logistics Management, 31(4), 247-265.
[74] Raschke, R. L. (2010). Process-based view of agility: The value contribution of IT and the effects on process outcomes. International Journal of Accounting Information Systems, 11(4), 297-313.
[75] Rdiouat, Y., Nakabi, N., Kahtani, K., & Semma, A. (2012). Towards a new approach of continuous process improvement based on CMMI and PMBOK. International Journal of Computer Science Issues (IJCSI).
[75b] Rockart, J. F. (1986). A Primer on Critical Success Factors" published in The Rise of Managerial Computing: The Best of the Center for Information Systems Research, edited with Christine V. Bullen.(Homewood, IL: Dow Jones-Irwin), 1981.OR
[76] Rosemann, M. (2001). Evaluating the management of enterprise systems with the Balanced Scorecard. Information Technology Evaluation Methods and Management, 171-184.
[77] Ross, J. W., Weill, P., & Robertson, D. (2006). Enterprise architecture as strategy: Creating a foundation for business execution. Harvard Business Press.
[78] Ross, J. W. (2003). Creating a Strategic IT Architecture Competency: Learning in Stages (No. CISR WP No. 335): MIT Sloan School.
[79] Sabherwal, R., Hirschheim, R., & Goles, T. (2001). The dynamics of alignment: Insights from a punctuated equilibrium model. Organization Science, 12(2), 179-197.
[80] Salmela, H., Tapaninainen, T., Baiyere, A., Hallanoro, M., & Galliers, R. (2015). IS Agility Research: An Assessment and Future Directions.
[81] Sambamurthy, V., Bharadwaj, A., Grover, V. (2003). Shaping agility through digital options: reconceptualizing the role of information technology in contemporary firms. MIS Quarterly, 27(2), 237-263.
[82] Sarhadi, M., & Millar, C. (2002). Defining a framework for information systems requirements for agile manufacturing. International Journal of Production Economics, 75(1), 57-68.
[83] Sengupta, K., Masini, A. (2008). IT agility: striking the right balance. Business Strategy Review, 19(2), 42-48.
[84] Sharifi, H., Colquhoun, G., Barclay, I., & Dann, Z. (2001). Agile manufacturing: a management and operational framework. Proceedings of the Institution of Mechanical Engineers, Part B: Journal of Engineering Manufacture, 215(6), 857-869.
[85] Sharifi, H., & Zhang, Z. (1998, July). Enabling practices assisting achievement of agile manufacturing. In Proceeding of the 6th IASTED International Conference (pp. 62-65).
[86] Sherehiy, B., Karwowski, W., & Layer, J. K. (2007). A review of enterprise agility: Concepts, frameworks, and attributes. International Journal of industrial ergonomics, 37(5), 445-460
[87] Shuiabi, E., Thomson, V., and Bhuiyan, N. (2005). Entropy as a measure of operational flexibility. European Journal of Operational Research, 165 (3), 696–707.
[88] Sidky, A. (2007). A structured approach to adopting agile practices: The agile adoption framework. Ph.D. thesis Virginia Polytechnic Institute and State University.
[89] Stewart, R. A., & Mohamed, S. (2001). Utilizing the balanced scorecard for IT/IS performance evaluation in construction. Construction innovation, 1(3), 147-163.
[90] Tallon, P. P. (2008). Inside the adaptive enterprise: an information technology capabilities perspective on business process agility. Information Technology & Management, 9(1), 21-36.
[91] Teoh, S. Y., & Cai, S. (2015). The Process of Strategic, Agile, Innovation Development: A Healthcare Systems Implementation Case Study. Journal of Global Information Management (JGIM), 23(3), 1-22.
[92] Tiwana, A., & Konsynski, B. (2010). Complementarities between organizational IT architecture and governance structure. Information Systems Research, 21(2), 288-304.
[93] Tsourveloudis, N.C., Valavanis, K.P., 2002. On the measurement of enterprise agility. Journal of Intelligent and Robotic Systems 33 (3), 329–342.
[94] Van Dick, R. & Ullrich, J. (2013). Identifikation und Commitment. In W. Sarges (Hrsg.). Managementdiagnostik (S. 349-354). Göttingen: Hogrefe
[95] Van Grembergen, W., & De Haes, S. (2005). Measuring and improving IT governance through the balanced scorecard. Information Systems Control Journal, 2(1), 35-42.
[96] Van Grembergen, W., Saull, R., & De Haes, S. (2003). Linking the IT balanced scorecard to the business objectives at a major Canadian financial group. Journal of Information Technology Case and Application Research, 5(1), 23-50.
[97] Van Grembergen, W., & Saull, R. (2001, January). Aligning business and information technology through the balanced scorecard at a major Canadian financial group: its status measured with an IT BSC maturity model. In System Sciences, 2001. Proceedings of the 34th Annual Hawaii International Conference on (pp. 10-pp). IEEE.

[98] Van Grembergen, W., & Van Bruggen, R. (1997, October). Measuring and improving corporate information technology through the balanced scorecard technique. In Proceedings of the Fourth European Conference on the Evaluation of Information technology (pp. 163-171).

[99] Van Grembergen, W. (2000) "The balanced scorecard and IT governance", Information Systems Control Journal (previously IS Audit & Control Journal), Volume 2, pp.40-43

[100] Van Can Grembergen, W. and Timmerman, D. (1998, Mai), Monitoring the IT process through the balanced score card", Proceedings of the 9th Information Resources Management (IRMA) International Conference, Boston, ,pp. 105-116

[101] Van Hoek, R. I., Harrison, A., Christopher, M. (2001). Measuring agile capabilities in the supply chain. International Journal of Operations & Production Management, 12(1/2), 126-147.

[102] Van Oosterhout, M. (2010). Business agility and information technology in service organizations (No. EPS-2010-198-LIS). Erasmus Research Institute of Management (ERIM).

[103] Van Weele, A.J. (2001) Purchasing and Supply Chain Management: Analysis, Planning & Practice(3rd revised edition). London, England: Thomson International.

[104] Sakthivel Aravindraj S. Vinodh , (2014),"Forty criteria based agility assessment using scoring approach in an Indian relays manufacturing organization", Journal of Engineering, Design and Technology, Vol. 12 Iss 4 pp. 507 – 518

[105] Vinodh, S., Madhyasta, U.R. and Praveen, T. (2012), "Scoring and multi-grade fuzzy assessment of agility in an Indian electric automotive car manufacturing organisation", International Journal of Production Research, Vol. 50 No. 3, pp. 647-660.

[105b] Waller, J., Ehmke, N. C., & Hasselbring, W. (2015). Including performance benchmarks into continuous integration to enable DevOps. ACM SIGSOFT Software Engineering Notes, 40(2), 1-4.

[106] Weill, P., Subramani, M., Broadbent, M. (2002). IT Infrastructure for Strategic Agility (No. CISR WP No. 329): Center for Information Systems Research, Sloan School of Management.

[107] Yang, S. L., & Li, T. F. (2002). Agility evaluation of mass customization product manufacturing. Journal of Materials Processing Technology, 129(1), 640-644.

[108] Yauch, C. A. (2011). Measuring agility as a performance outcome. Journal of Manufacturing Technology Management, 22(3), 384-404.

[109] Yusuf, Y. Y., Sarhadi, M., & Gunasekaran, A. (1999). Agile manufacturing:: The drivers, concepts and attributes. International Journal of production economics, 62(1), 33-43.

[110] Zhang, Z., & Sharifi, H. (2000). A methodology for achieving agility in manufacturing organisations. International Journal of Operations & Production Management, 20(4), 496-513.